\documentclass[pra,showpacs,twocolumn,superscriptaddress]{revtex4}

\newcommand{\ket}[1]{|{#1}\rangle}
\newcommand{\bra}[1]{\langle{#1}|}

\newcommand{\be}{\begin{equation}}
\newcommand{\ee}{\end{equation}}

\usepackage{amsfonts}
\usepackage{amsmath}
\usepackage{graphicx}
\usepackage{amssymb}
\usepackage{amsmath}
\usepackage{amssymb}
\usepackage{graphicx}
\usepackage{lscape}
\usepackage{color}
\usepackage{epstopdf}

\begin{document}
\title{Quantum collision theory in flat bands}

\author{Manuel Valiente}
\affiliation{SUPA, Institute of Photonics and Quantum Sciences, Heriot-Watt University, Edinburgh EH14 4AS, United Kingdom}
\affiliation{Kavli Institute for Theoretical Physics, University of California, Santa Barbara, Santa Barbara, CA 93106}
\author{Nikolaj Thomas Zinner}
\affiliation{Department of Physics \& Astronomy, Aarhus University, 8000 Aarhus C, Denmark}
\affiliation{Kavli Institute for Theoretical Physics, University of California, Santa Barbara, Santa Barbara, CA 93106}

\begin{abstract}
We consider quantum scattering of particles in media exhibiting strong dispersion degeneracy. In particular, we study flat-banded lattices and linearly dispersed energy bands. The former constitute a prime example of single-particle frustration while the latter show degeneracy at the few- and many-particle level. We investigate both impurity and two-body scattering and show that, quite generally, scattering does not occur, which we relate to the fact that transition matrices vanish on the energy shell. We prove that scattering is instead replaced by projections onto band-projected eigenstates of the interaction potential. We then use the general results to obtain localised flat band states that are eigenstates of impurity potentials with vanishing eigenvalues in one-dimensional flat bands and study the particular case of a sawtooth lattice. We also uncover the relation between certain solutions of one-dimensional systems that have been categorised as "strange", and the scattering states in linearly dispersed continuum systems.       
 
\end{abstract}
\pacs{
}
\maketitle
\section{Introduction}
The behavior of quantum systems is to a large extent controlled 
by the competition between the kinetic energy and the interaction 
energy. The former tends to drive a system towards delocalisation 
in order to minimise the spatial variation of quantum states, 
while the latter tends to drive the system towards localisation 
in the minima of the interaction potential. This is the case
irrespective of whether one is interested in few- or many-body
systems, and whether one studies particles on a lattice 
or in the continuum. 
For instance, in few-body physics bound states are 
more likely in lower spatial dimension due to reduced kinetic 
energy \cite{blume2012,zinnerjensen2013,zinner2014,naidon2016}.
For many-body systems, there is also a strong dependence on the 
spatial dimension when considering collective quantum states such 
as superconductors and condensates as the coherence of many-body 
systems becomes relatively weaker in lower dimension \cite{bkt1,bkt2,bkt3}.

Another path to the study of kinetic versus interaction energy
in quantum systems is via changes to the dispersion relation, i.e. 
to the particular form that the kinetic energy takes as a function 
of the momentum. By changing the dispersion relation from the 
usual quadratic form one may influence the competition of effects
and thus explore a richer variety of quantum behaviour. Here we 
are interested in few-body physics with non-standard dispersion 
relations. The first step towards an understanding of such systems
is the determination of the scattering properties of particles, 
i.e. single particle scattering off an impurity that is represented 
by a given potential and two particles scattering off each other. 

An interesting and important case is the limit where the 
dispersion is essentially flat in the sense that the single particle
energy is independent of the momentum or quasi-momentum of the particle. In the context
of lattices, this is called a flat band. The lack of kinetic energy
means that interactions are decisive, and one may expect to see strong
localization in such systems. However, superfluidity and kinetic behaviour can be induced by interactions
\cite{Peotta1,Peotta2,Peotta3,Julien0,Gershenson}.
 This sort of band can be achieved
in many different lattices including the dice lattice \cite{sutherland1986,Julien1,Julien2,Julien3}, 
stub lattice (1D Lieb lattice) or diamond lattices \cite{hyrkas2013},
the sawtooth lattice \cite{zhang2015,phillips2015}, Kagome lattices \cite{santos2004}, 
and Lieb lattices \cite{yang2016}. For a theoretical discussion on 
how to construct general flat band lattices see Ref.~\cite{morales2016}.
Flat bands have been considered in many-body physics of the Hubbard 
model as a driver
for ferromagnetism on so-called line graphs \cite{mielke1991}
of which the Kagome lattice is one example \cite{mielke1992}, 
and more generally if the flat band has more states than there
are electrons in the system \cite{mielke1993}.
Lattices that contain flat bands have been realized experimentally
using a diverse variety of physical systems including 
stub lattices with optical cavities \cite{baboux2016}, 
sawtooth \cite{weimann2016}, Kagome \cite{zong2016}, and 
Lieb lattices \cite{vicencio2015} with
photonic lattices, Kagome \cite{jo2012} and 
Lieb \cite{taie2015} lattices with cold atoms, 
and most recently the Lieb lattice realization 
using electronic surface states on an appropriately 
doped surface has been reported \cite{slot2016}.

Flat bands are a very interesting case of the general 
phenomenon of non-trivial dispersion relations, and systems
that display such behavior are attracting much attention at 
the moment. A particularly prominent example is graphene, a
two-dimensional material with a linear dispersion around the 
Fermi energy, i.e. a behavior expected of particles in the 
relativistic regime (or massless particles). This feature 
can be generalized to three-dimensional materials with linear
dispersions which are typically called either Dirac or Weyl 
semi-metals depending on whether time-reversal or parity symmetry 
is broken or not. In fact, a new subfield has emerged studying
these so-called Dirac materials \cite{wehling2014}. These systems are
very interesting also from a few-body
point of view due to their non-trivial scattering properties 
\cite{sabio2010,huang2013,gaul2014,jiang2016}. We note that 
truly flat bands are also of great interest in the context
of graphene \cite{lothman2016,chen2016}.

Within the field 
of cold atoms, a recent direction of great interest is the 
production of artificial gauge potentials \cite{dalibard2011}. 
In particular, different kinds of spin-orbit coupling known 
mostly from condensed-matter contexts, Rashba/Dresselhaus 
spin-orbit coupling, have been explored extensively both
theoretically and experimentally \cite{zhai2012,galitski2013}. 
In the few-body community there has also been a flurry of 
activity in spin-orbit coupled systems with its non-trivial 
dispersion. In particular, the scattering properties have
been discussed recently \cite{mundo2013a,mundo2013b,wang2015,zhou2015,guan2016,wang2016}
as well as the effects of spin-orbit coupling on systems that 
are in a harmonic trap \cite{achilleos2013,marchukov2013,guan2014,yin2014,marchukov2014,guan2015}. 
Furthermore, the three-body physics of these systems has also 
been found to be very rich \cite{shi2014,cui2014,shi2015}. 

In this paper we study potential scattering and two-body scattering for particles that either live on a lattice with a flat single-particle dispersion, or that are in a continuum system with a purely linear dispersion. We find in general that there is no scattering, strictly speaking, in these systems, and scattering states can be written as a projection of the incident Bloch waves onto the null subspace of the interaction potential, thereby generalising the continuum results for intrabranch collisions in a Luttinger liquid \cite{Phillips}. We also construct localised eigenstates for these systems in one dimension and apply our general results to a particular flat-banded system -- the so-called sawtooth lattice.

\section{Potential scattering}\label{potentialscattering}
We begin by considering the simplest non-trivial scenario consisting of a single particle colliding with a short-range potential in an otherwise periodic lattice structure. We denote the non-interacting Hamiltonian by $\hat{H}_0$ and the ``impurity'' potential by $\hat{V}$. We further assume, without loss of generality, that the system possesses a finite number of bands labeled by $\sigma=1,\ldots,n$, one of them being flat, namely $\sigma_n$, while the other $n-1$ bands, i.e. $\sigma_1,\sigma_2,\ldots,\sigma_{n-1}$, are in principle dispersive. We denote the different energy dispersions by $E_{\sigma_j}(\mathbf{k})$, with $\mathbf{k}$ in the first Brillouin zone (1BZ), while their corresponding eigenstates (Bloch waves), are denoted by $\ket{\sigma_j \mathbf{k}}$, such that
\begin{equation}
\hat{H}_0\ket{\sigma \mathbf{k}}=E_{\sigma}(\mathbf{k})\ket{\sigma \mathbf{k}}.
\end{equation}
We further assume that the flat band is gapped, that is, $E_{\sigma_n}(\mathbf{k})\equiv E_{\sigma_n} \ne E_{\sigma_j}(\mathbf{k})$ for $j=1,\ldots,n-1$ and $\forall \mathbf{k}\in \mathrm{1BZ}$. 

In order to evaluate the collisional properties of the system, we need to solve the Lippmann-Schwinger equation for the T-matrix at (complex) energy $z$, which reads 
\begin{equation}
\hat{T}(z)=\hat{V}+\hat{V}\hat{G}^{(0)}(z)\hat{T}(z).
\end{equation}
Above, $G^{(0)}(z)\equiv (z-\hat{H}_0)^{-1}$ is the non-interacting Green's function at energy $z$. We work directly in the infinite volume limit, and in the quasi-momentum representation. We denote the volume of the first Brillouin zone by $\Omega$, and normalize the single-particle states such that $\langle\sigma' \mathbf{k}'|\sigma \mathbf{k}\rangle=\Omega\delta_{\sigma,\sigma'}\delta(\mathbf{k}'-\mathbf{k})$. With these definitions, a resolution of the identity can be written as
\begin{equation}
\hat{1}=\sum_{\tau=1}^{n}\frac{1}{\Omega}\int_{\mathrm{1BZ}}d\mathbf{q}\ket{\tau \mathbf{q}}\bra{\tau \mathbf{q}}.\label{resolution}
\end{equation}  
The equations satisfied by the elements of the T-matrix are then obtained by inserting Eq. (\ref{resolution}) into the Lippmann-Schwinger equation, which yields
\begin{align}
&\bra{\sigma'\mathbf{k}'}\hat{T}(z)\ket{\sigma \mathbf{k}}=\bra{\sigma' \mathbf{k}'}\hat{V}\ket{\sigma \mathbf{k}}\nonumber\\
&+\sum_{\tau=1}^{n}\frac{1}{\Omega}\int_{\mathrm{1BZ}}d\mathbf{q}\frac{\bra{\sigma'\mathbf{k}'}\hat{V}\ket{\tau \mathbf{q}}\bra{\tau \mathbf{q}}\hat{T}(z)\ket{\sigma \mathbf{k}}}{z-E_{\tau}(\mathbf{q})}.\label{longLS}
\end{align}
Until here, we have simply written down the Lippmann-Schwinger equation of potential scattering for a general multiband system. We now proceed to studying scattering of a particle with incident quasi-momentum $\mathbf{k}$ and incident band index $\sigma_n$, that is, on the flat band. To this end, we go on the energy shell by setting $z\to z_0 = E_{\sigma_n}+i\eta$, with $\eta$ an infinitesimal imaginary part of the energy that is taken to 0 ($\eta \to 0^{+}$) at the end of the calculation. Separating the contribution $\tau=\sigma_n$ from the rest of the sum on the second line of Eq. (\ref{longLS}), we can re-write it as
\begin{align}   
&\bra{\sigma'\mathbf{k}'}\hat{T}(z_0)\ket{\sigma \mathbf{k}}=\bra{\sigma' \mathbf{k}'}\hat{V}\ket{\sigma \mathbf{k}}\nonumber\\
&+\sum_{\tau\ne \sigma_n}\frac{1}{\Omega}\int_{\mathrm{1BZ}}d\mathbf{q}\frac{\bra{\sigma'\mathbf{k}'}\hat{V}\ket{\tau \mathbf{q}}\bra{\tau \mathbf{q}}\hat{T}(z_0)\ket{\sigma \mathbf{k}}}{E_{\sigma_n}-E_{\tau}(\mathbf{q})}\nonumber \\
&+\frac{1}{\Omega}\int_{\mathrm{1BZ}}d\mathbf{q}\frac{\bra{\sigma'\mathbf{k}'}\hat{V}\ket{\sigma_n \mathbf{q}}\bra{\sigma_n \mathbf{q}}\hat{T}(z_0)\ket{\sigma \mathbf{k}}}{i\eta}.\label{onshellT}
\end{align}
Notice that on the second line of the above equation we have used our assumption that the flat-band is gapped. It is now immediate to realize that, for Eq. (\ref{onshellT}) to have a non-trivial solution, the on-shell T-matrix elements of the form $\bra{\sigma_n\mathbf{q}}\hat{T}(z_0)\ket{\sigma \mathbf{k}}$ must vanish as $\sim \eta$ for $\eta \to 0^+$. In order to get rid of the seemingly divergent term in Eq. (\ref{onshellT}), we define 
\begin{equation}
\bra{\sigma_n\mathbf{q}}\hat{T}(z_0)\ket{\sigma \mathbf{k}}\equiv i\eta\bra{\sigma_n\mathbf{q}}\hat{t}(z_0)\ket{\sigma \mathbf{k}},\label{definet}
\end{equation}
and therefore Eq. (\ref{onshellT}) reads
\begin{align}   
&\bra{\sigma'\mathbf{k}'}\hat{T}(z_0)\ket{\sigma \mathbf{k}}=\bra{\sigma' \mathbf{k}'}\hat{V}\ket{\sigma \mathbf{k}}\nonumber\\
&+\sum_{\tau\ne \sigma_n}\frac{1}{\Omega}\int_{\mathrm{1BZ}}d\mathbf{q}\frac{\bra{\sigma'\mathbf{k}'}\hat{V}\ket{\tau \mathbf{q}}\bra{\tau \mathbf{q}}\hat{T}(z_0)\ket{\sigma \mathbf{k}}}{E_{\sigma_n}-E_{\tau}(\mathbf{q})}\nonumber \\
&+\frac{1}{\Omega}\int_{\mathrm{1BZ}}d\mathbf{q}\bra{\sigma'\mathbf{k}'}\hat{V}\ket{\sigma_n \mathbf{q}}\bra{\sigma_n \mathbf{q}}\hat{t}(z_0)\ket{\sigma \mathbf{k}}.\label{onshellT2}
\end{align} 
Consider now, in Eq. (\ref{onshellT2}), the matrix elements with $\sigma'=\sigma_n$. After using the definition (\ref{definet}), and taking the limit $\eta\to 0^+$, we obtain
\begin{align}
&-\bra{\sigma_n\mathbf{k}'}\hat{V}\ket{\sigma \mathbf{k}}=\nonumber\\
&\sum_{\tau\ne \sigma_n}\frac{1}{\Omega}\int_{\mathrm{1BZ}}d\mathbf{q}\frac{\bra{\sigma_n\mathbf{k}'}\hat{V}\ket{\tau \mathbf{q}}\bra{\tau \mathbf{q}}\hat{T}(z_0)\ket{\sigma \mathbf{k}}}{E_{\sigma_n}-E_{\tau}(\mathbf{q})}\nonumber \\
&+\frac{1}{\Omega}\int_{\mathrm{1BZ}}d\mathbf{q}\bra{\sigma_n\mathbf{k}'}\hat{V}\ket{\sigma_n \mathbf{q}}\bra{\sigma_n \mathbf{q}}\hat{t}(z_0)\ket{\sigma \mathbf{k}}.\label{onshellT3}
\end{align} 
The final system of equations that must be solved for the on-shell T-matrix is therefore given by Eqs. (\ref{onshellT2}) for $\sigma'\ne \sigma_n$, and Eq. (\ref{onshellT3}) when $\sigma'=\sigma_n$. 

We now have all that is necessary to study collisions in flat bands. However, just with the Lippmann-Schwinger set of equations (\ref{onshellT2}) and (\ref{onshellT3}) the physics is not very transparent. We can easily see, from Eqs.~(\ref{onshellT3}) and (\ref{definet}), that $\bra{\sigma_n \mathbf{q}}\hat{t}(z_0)\ket{\sigma_n \mathbf{k}}$ satisfies 
\begin{equation}
-\bra{\sigma_n\mathbf{k}'}\hat{V}\ket{\sigma_n\mathbf{k}}= \frac{1}{\Omega}\int_{\mathrm{1BZ}}d\mathbf{q}\bra{\sigma_n\mathbf{k}'}\hat{V}\ket{\sigma_n\mathbf{q}}\bra{\sigma_n\mathbf{q}}\hat{t}(z_0)\ket{\sigma_n\mathbf{k}}.\label{equref}
\end{equation}
To prove the above, we only needed to use that $\bra{\tau \mathbf{q}}\hat{T}(z_0)\ket{\sigma_n\mathbf{k}}=0$, which is granted by Eq.~(\ref{definet}).
In the projected flat band subspace, Eq.~(\ref{equref}) can be rewritten as 
\begin{equation}
-\hat{V} = \hat{V}\hat{t}(z_0).\label{VVt}
\end{equation}
We diagonalise $\hat{V}$ in the flat band subspace in order to solve Eq.~(\ref{VVt}). We choose a unitary operator $\hat{U}$  that diagonalizes $\hat{V}$. The corresponding orthogonal set of eigenfunctions are $\ket{\sigma_n\alpha_{\mathbf{k}}}$ and their eigenvalues are denoted by $V(\alpha_{\mathbf{k}})$. This means
\begin{equation}
\ket{\sigma_n\alpha_{\mathbf{k}}}=\hat{U}\ket{\sigma_n\mathbf{k}}.
\end{equation}
Eq. (\ref{VVt}) becomes
\begin{equation}  
-\Omega V(\alpha_{\mathbf{q}})\delta(\mathbf{q}-\mathbf{q})=V(\alpha_{\mathbf{q}'})\bra{\sigma_n\alpha_{\mathbf{q}'}}\hat{t}(z_0)\ket{\sigma_n\alpha_{\mathbf{q}}}.\label{VVt2}
\end{equation}
For a short-range potential $\hat{V}$, most of its eigenvalues vanish. We define the set of vanishing eigenvalues as
\begin{equation}
I_0=\{ \mathbf{q}\in \mathrm{1BZ}:V(\alpha_{\mathbf{q}})=0 \},
\end{equation}
and $I_V\equiv \mathrm{1BZ}-I_0$. For $\mathbf{q}$ and $\mathbf{q}'$ in the set $I_0$, the matrix elements $\bra{\sigma_n\alpha_{\mathbf{q}'}}\hat{t}(z_0)\ket{\sigma_n \alpha_{\mathbf{q}}}$ are arbitrary, as is seen from Eq.~(\ref{VVt2}). If $\mathbf{q}'$ is in $I_V$ instead, then  $\bra{\sigma_n\alpha_{\mathbf{q}'}}\hat{t}(z_0)\ket{\sigma_n \alpha_{\mathbf{q}}}=0$. For $\mathbf{q}$ in $I_V$, if $\mathbf{q}\ne \mathbf{q}'$ we have that $\bra{\sigma_n\alpha_{\mathbf{q}'}}\hat{t}(z_0)\ket{\sigma_n \alpha_{\mathbf{q}}}$ is arbitrary, while if $\mathbf{q}=\mathbf{q}'$ we have $\bra{\sigma_n\alpha_{\mathbf{q}'}}\hat{t}(z_0)\ket{\sigma_n \alpha_{\mathbf{q}}}=-2\pi$.


We now shall calculate the scattering states. We split the scattering state $\ket{\psi}$ into incident $\ket{\psi_0}$ and scattered $\ket{\psi_s}$ waves, as
\begin{equation}
\ket{\psi}=\ket{\psi_0}+\ket{\psi_s}.
\end{equation}
The relation between scattered and incident waves in terms of the T-matrix is given by
\begin{equation}
\ket{\psi_s}=\hat{G}^{(0)}(z_0)\hat{T}(z_0)\ket{\psi_0}.
\end{equation}
The incident state is nothing but a flat band state, $\ket{\psi_0}=\ket{\sigma_n\mathbf{k}}$. The  scattered wave has the form
\begin{align}
\langle \sigma_n \mathbf{k}'|\psi_s\rangle &= \bra{\sigma_n\mathbf{k}'}\hat{G}^{(0)}(z_0)\hat{T}(z_0)\ket{\sigma_n\mathbf{k}}\nonumber \\
&=\frac{1}{i\eta}\bra{\sigma_n\mathbf{k}'}\hat{T}(z_0)\ket{\sigma_n \mathbf{k}}\nonumber\\
&=\bra{\sigma_n \mathbf{k}'}\hat{t}(z_0)\ket{\sigma_n \mathbf{k}},
\end{align}
where in the last step we have used Eq. (\ref{definet}). We now set all the arbitrary matrix elements discussed above to a constant value $\gamma$. Using the T-matrix calculated above, and after some algebraic manipulations, we find that the total scattering wave function is given by
\begin{align}
&\ket{\psi}= \ket{\sigma_n\mathbf{k}}+\int_{\mathrm{1BZ}}\frac{d\mathbf{k}'}{\Omega}\ket{\sigma_n\mathbf{k}'}\left[ \gamma \int_{I_0}\frac{d\mathbf{q}'}{\Omega}\langle \sigma_n\mathbf{k}'|\sigma_n\alpha_{\mathbf{q}'}\rangle \right. \nonumber \\
&\left. \times \int_{\mathrm{1BZ}}\frac{d\mathbf{q}'}{\Omega}\langle \sigma_n\alpha_{\mathbf{q}}|\sigma_n\mathbf{k}\rangle -\int_{I_V}\frac{d\mathbf{q}}{\Omega}\langle \sigma_n\mathbf{k}'|\sigma_n\alpha_{\mathbf{q}}\rangle \langle \sigma_n\alpha_{\mathbf{q}}|\sigma_n\mathbf{k}\rangle\right] .\label{fullwf}
\end{align} 
We now make the simple choice $\gamma=0$. This does not reduce the generality of 
our discussion as any other choice of $\gamma$ would merely result in a different 
projection operator within the flat band and thus would yield no new solutions.
Together with the resolution of the identity,  
\begin{equation}
\hat{1}_{\sigma_n}=\int_{\mathrm{1BZ}}\frac{d\mathbf{k}'}{\Omega}\ket{\sigma_n\mathbf{k}'}\bra{\sigma_n\mathbf{k}'},
\end{equation}
and using that
\begin{equation}
\hat{1}_{\sigma_n}-\hat{P}_{I_0}=\int_{I_V}\frac{d\mathbf{q}}{\Omega}\ket{\sigma_n\alpha_{\mathbf{q}}}\bra{\sigma_n\alpha_{\mathbf{q}}},
\end{equation}
where $\hat{P}_{I_0}$ is the projector onto the eigenspace of $\hat{V}$ associated with eigenvalue zero (on the projected flat band subspace), we now obtain
\begin{equation}
\ket{\psi_{\gamma=0}}=\hat{P}_{I_0}\ket{\sigma_n \mathbf{k}}.\label{omega0}
\end{equation}
We see that the scattering states obtained in this way are constructed by taking a non-interacting Bloch wave in the corresponding flat band $\sigma_n$, and projecting it onto the subspace of zero eigenvalues of the potential $\hat{V}$.

\subsection{Localised states}
We turn our attention now to localised states in flat-banded one-dimensional lattices. These occur generally in highly frustrated systems, and correspond to a change of basis from usual Bloch waves. The physics of localised states is easy to grasp: since all Bloch waves share the same energy, any arbitrary superposition thereof is an eigenstate, too. If the flat band has a non-vanishing density of states~\footnote{If $V$ is the lattice volume, and $M$ the number of states in the flat band, then $M/V\ne 0$ as $V\to \infty$.}, we can integrate over all flat band modes with particular weights in order to obtain a localised eigenstate. It is simple to visualize this scenario by considering a toy model. Assume that a particle in a one dimensional lattice has a single dispersionless ``band'', and the eigenstates are plane waves with momentum $k$. The corresponding flat band can be taken to have zero energy ($E(k)=0$). The following states are also eigenstates     
\begin{equation}
\delta_{x,x_0}=\frac{1}{2\pi}\int_{-\pi}^{\pi}dk \exp(ikx) \exp(-ikx_0),
\end{equation}
and are obviously localized. This model, as unphysical as it seems, already contains all we need to know about localized flat band states in more realistic models, as we will see.

We begin by considering a general one dimensional system with a flat band, labeled by $\sigma_n$. We introduce a probe impurity modeled by a potential $\hat{V}$. As is clear from Eq. (\ref{omega0}), scattering states on the flat band satisfy $\hat{V}\ket{\psi}=0$. As before, we denote the Bloch waves by $\ket{\sigma_nk}$, with $k\in \mathrm{1BZ}$, and the Brillouin zone volume by $\Omega$. We perform a change of basis as follows
\begin{equation}
\ket{\sigma_n y} = \frac{2}{\Omega}\int_{-\Omega/2}^{\Omega/2}dk y(k)\ket{\sigma_nk},\label{yexpansion}
\end{equation}
The condition $\ket{\sigma_n y}\in I_0$ is satisfied if, for all $q$,
\begin{equation}
\int_{-\Omega/2}^{\Omega/2}dk y(k) \bra{\sigma_nq}\hat{V}\ket{\sigma_nk}=0.\label{yeq1}
\end{equation}
The matrix elements of $\hat{V}$ are, by construction, given by
\begin{equation}
\bra{\sigma_nq}\hat{V}\ket{\sigma_nk}=A(q) A(k) \sum_{x=-\infty}^{\infty} \left[f_x(q)\right]^*f_x(k)e^{i(k-q)x},\label{fullpot}
\end{equation}
where $A(k)$ is a normalization constant (which we have chosen to be real and positive), while $f_x(k)$ are $x$-dependent functions. We are free to choose probe impurity potentials consisting of zero-range potentials acting on a given site $x_0$, which we label by $\hat{V}_{x_0}^{(0)}$, and have the form (see Eq. (\ref{fullpot}))
\begin{equation}
\bra{\sigma_nq}\hat{V}_{x_0}^{(0)}\ket{\sigma_nk}\propto A(k) f_{x_0}(k) e^{ikx_0},
\end{equation}
where we have dropped the irrelevant factors that depend on $q$. With these considerations, if $f_{x_0}(k)$ has no nodes, Eq. (\ref{yeq1}) for the zero range potential $\hat{V}_x^{(0)}$ has solutions of the form
\begin{equation}
y_{\ell}(k)=e^{-ikx_0}\left[A(k)f_{x_0}(k)\right]^{-1}e^{2\pi i \ell k/\Omega}, \hspace{0.1cm} \ell \in \mathbb{Z}-\{0\}.\label{yell}
\end{equation}      
If, on the other hand, $f_{x_0}(k_*)=0$ for some $k_*$, then obviously $\hat{V}_{x_0}^{(0)}\ket{\sigma_nk_*}=0$ and one solution is given by $y(k)\propto \delta(k-k_*)$. Other solutions can be constructed by eliminating this state from the expansion.


\subsection{Example: Sawtooth Lattice}
We consider an impurity located at $x=0$ in the so-called sawtooth lattice in one dimension. The single-particle Schr{\"o}dinger equation is given by \cite{phillips2015}
\begin{equation}
t\sum_{\mu=\pm 1}\left[\sqrt{2}\psi(x+\mu)+\frac{(1+(-1)^x)}{2}\psi(x+2\mu)\right]=E\psi(x),
\end{equation}
where $x$ are integer lattice sites, and we have set the lattice spacing $d=1$. This system has two energy bands, one of which is flat while the other one is dispersive. These are given by
\begin{align}
E_0(k)&=-2t\\
E_1(k)&=2t(1+\cos 2k).
\end{align}
The eigenfunctions in the flat band are given by Bloch's theorem, i.e. $\psi_k(x)=\phi_k(x)\exp(ikx)$, where $\phi_k(x+2)=\phi_k(x)$. The two relevant values of the Bloch functions $\phi_k(0)$ and $\phi_k(1)$ are given by
\begin{align}
\phi_k(0)&=\frac{1}{\sqrt{1+2\cos^2 k}},\\
\phi_k(1)&=-\sqrt{2}\cos (k) \phi_k(0).
\end{align}
We see that the only points at which the "issue" with the nodes happens is at $k=\pm \pi/2$. In that case, $\phi_k(0)$ is a constant while $\phi_k(1)$ vanishes. That is, the eigenfunctions in the flat band with $k=\pm \pi/2$ vanish at all odd sites, and those do not see an impurity located at an odd site. For an impurity located at an even site, such as the case we are considering, the eigenfunctions are all non-zero at even sites and there is no such issue with the nodes. 

It is now straightforward to calculate eigenstates of the impurity potential with zero eigenvalue. We  obtain
\begin{equation}
y_{\ell}(k)=\sqrt{1+2\cos^2 k} e^{2i\ell k}, \ell \in \mathbb{Z}-\{0\}.
\end{equation}
In the position representation, these eigenfunctions $\alpha_{\ell}(x)$ get the form 
\begin{align}
\alpha_{\ell}(x)&=\int_{-\pi/2}^{\pi/2}dke^{i(2\ell + x)k}, x \hspace{0.1cm} \mathrm{even}\\
\alpha_{\ell}(x)&=-\sqrt{2}\int_{-\pi/2}^{\pi/2}dk \cos k e^{i(2\ell +x)k}, x \hspace{0.1cm} \mathrm{odd}.
\end{align}
Interestingly, the wave functions as constructed above immediately become, as one would na\"ively expect, completely localised, and form the so-called "V-states" of the sawtooth lattice \cite{phillips2015},
\begin{equation}
\alpha_{\ell}(x)=\pi \delta_{x,-2\ell}-\frac{\pi}{\sqrt{2}}\left(\delta_{x,-2\ell+1}+\delta_{x,-2\ell-1}\right),
\end{equation}
which exclude the $\ell=0$ term -- the only state that overlaps with the impurity. If we add the $\ell=0$ state to the above set of eigenstate (which can be obtained by moving the impurity to $x=2$), we actually recover the localised flat-band basis \cite{huber2010}.

\section{Two-body problems}
We consider now two or more particles interacting with each other via a two-body potential $\hat{V}$. We study two distinct scenarios below. Firstly, we generalise, very briefly, the result of the previous section on potential scattering for two particles in single-particle flat bands. We then move on to study collision theory for one-dimensional systems with linear dispersions, which is in many ways analogous to particles in flat bands. 

\subsection{Flat bands}
We first consider the two-body problem in a system admitting a flat band. We use the same notations and definitions of Section \ref{potentialscattering}. The difference here is the fact that the potential $\hat{V}$ is here a two-body interaction. We further assume that the interaction conserves centre of mass quasimomentum $\mathbf{K}=\mathbf{k}_1+\mathbf{k}_2$, where $\mathbf{k}_i$ is the incident quasi-momentum of the Bloch wave corresponding to particle $i$ ($i=1,2$). The two-body scattering states are calculated in the same way as in Section \ref{potentialscattering} and simply read (see Eq. (\ref{omega0}))
\begin{equation}
\ket{\psi}=\hat{P}_{I_0}\ket{\sigma_n\mathbf{k}_1,\sigma_n\mathbf{k}_2}.\label{omega02}
\end{equation}
Above, the set $I_0$ is just the degenerate eigenspace of $\hat{V}$ associated with eigenvalue zero on the projected flat band subspace. We can generalize the two-body result to any number of particles, provided we are at zero density or, more generally, at densities below the flat band critical density $\nu_c$ for which $\hat{V}$ would have no vanishing eigenvalues in the projected subspace. 

\subsection{Linear dispersions}
In one dimensional many-body systems, dispersion relations for collective excitations are typically linearized around the Fermi points $\pm k_F$. This lead to Tomonaga's \cite{Tomonaga} and Luttinger's models \cite{Luttinger}. While Tomonaga solved the many-body problem at finite densities correctly, Luttinger found a solution to his model which is only correct if the Fermi sea is not filled. Note that in Luttinger's model the filled Fermi sea corresponds to {\it infinite} density, and Luttinger's solutions are actually correct at finite densities, although they do not give the ground state \footnote{Note also that there is no ground state without momentum cutoffs as one may then populate states with infinitely large negative momentum.}, at finite densities. The infinite density limit of Luttinger's model was later solved by Mattis and Lieb in Ref. \cite{MattisLieb}. Mattis and Sutherland gave later on further solutions for one-dimensional linearly dispersed fermions, which they called ``strange'' \cite{MattisSutherland}. These correspond to localized wave functions that minimize the potential energy. We show here, from a scattering-theoretical point of view, that these ``strange'' solutions are in one-to-one correspondence with flat-band scattering states. 

For simplicity, we consider a linear, single branch problem in continuous space. The single-particle dispersion reads
\begin{equation}
E(\mathbf{k})=\hbar v_0 k,
\end{equation}
and we work first without momentum cutoffs $\pm \Lambda$, that is, we take the limit $\Lambda \to \infty$ before we try to find the stationary scattering states. The two-body interaction $\hat{V}$ is assumed to conserve total momentum. Then, any two particles with total momentum $K=k_1+k_2$ have the same energy $E(K)=\hbar v_0 K$ regardless of their relative momentum $k=(k_1-k_2)/2$. Therefore, for each $K$ we effectively have a potential scattering problem in a flat band. If the interaction has vanishing eigenvalues (that is, if it has nodes or has finite range), then scattering states exist and they have the form of Eq. (\ref{omega02}). If we now assume that $\hat{V}$ has finite range, that is, $V(x_1-x_2)=0$ for $|x_1-x_2|>R$, then there are infinitely many scattering states. Since these all share the same energy, for fixed $K$, an arbitrary superposition of them also has the same energy. Hence, if $\ket{\psi_{k_1,K-k_1}}=\hat{P}_{I_0}\ket{k_1,K-k_1}$, the following vector $\ket{\phi_K}$ is also an eigenstate
\begin{equation}
\ket{\phi_K}=\int_{-\infty}^{\infty}dk_1 \phi_K(k_1)\ket{\psi_{k_1,K-k_1}}.
\end{equation}
In particular, we can choose a completely co-localized relative wave function, and have the simple form
\begin{equation}
\ket{\phi_K}=\ket{K}\ket{x},
\end{equation}
for all $x$ for which $V(x)=0$. Obviously, all eigenstates, including those that are not scattering states, are eigenstates of the potential, and an orthonormal basis of them is given by the position eigenstates. These are what Mattis and Sutherland called ``strange'' solutions in Ref. \cite{MattisSutherland}. One of the reasons for this name was simply that they do not reduce to Slater determinants (for fermions) constructed with plane waves in the limit of zero interaction. As we see now from our perspective, this is not a problem, since, because of degeneracy, plane waves can be trivially constructed as superpositions of position eigenstates. The other, more physical reason, is that from these solutions one cannot construct the ground state of the infinitely filled Fermi sea or, equivalently for finite numbers of particles, the ground state with a momentum cutoff at $-\Lambda$. However, this ground state, which is easily obtained by means of bosonization, is also an eigenstate of the interaction. To see this, let us write down the second-quantized Hamiltonian for fermions in the momentum representation with cutoffs at $k=\pm\Lambda$,
\begin{equation}
H=\hbar v_0\sum_{k= -\Lambda}^{\Lambda}kc_k^{\dagger}c_k+\frac{1}{2L}\sum_{k,k',q}V(q)c_{k+q}^{\dagger}c_{k'-q}^{\dagger}c_{k'}c_k,
\end{equation}
where $c_k$ ($c_k^{\dagger}$) annihilates (creates) a fermion with momentum $k$. The interaction part of the Hamiltonian conserves total momentum. Clearly, for $N$ fermions, the state of lowest total momentum $K=k_1+\ldots+k_N$ corresponds to the Fermi sea $\ket{F}=\prod_{i=1}^Nc_{k_i}^{\dagger}\ket{0}$, and is unique. Since $[\hat{K},\hat{H}]=0$, we know that $\hat{H}\ket{F}$ is also an eigenstates of $\hat{K}$ with momentum $K$. But by the uniqueness of $\ket{F}$, they must be equal up to a multiplicative factor. Thus $\ket{F}$ is an eigenstate of $\hat{H}$. Using the fact that $\ket{F}$ is an eigenstate of the non-interacting Hamiltonian implies that $\ket{F}$ is an eigenstate of the interaction.


\section{Discussion}
We have considered what standard scattering theory has to say about collisions in highly-degenerate, or flat bands. We have found that, while the on-shell T-matrix vanishes in general linearly with energy, i.e. $\hat{T}_{\mathrm{on-shell}}\sim i\eta \hat{t}$ for $\eta\to 0^+$, it has a non-trivial structure given by $\hat{t}$ which can be used to extract scattering states. We have also found that, for incident particles in the flat band, interband transitions are forbidden, and scattering states only exist if the interaction potential has at least one vanishing eigenvalue. Due to the extreme energy degeneracy, scattering states are not uniquely defined, and we have shown how a particular choice for certain arbitrary T-matrix elements yields physically intuitive stationary scattering states. These are simply given by the projection of the incident Bloch waves onto the null subspace of the interaction potential. Using this fact, we have shown how to construct localised states in flat bands that are scattering states off impurity potentials in one dimension, and particularised to a sawtooth lattice. We have also considered linearly dispersed continuum systems and showed how our scattering states are related to the strange solutions of one-dimensional quantum field theories found by Sutherland and Mattis \cite{MattisSutherland}.

\acknowledgments
The authors are grateful for the support of the Kavli Institute for 
Theoretical Physics in Santa Barbara for hosting the program 
'Universality in Few-Body Systems'. We thank the participants of this 
program for interesting discussions. One of the topical questions of 
interest at the program was
'What about universal properties of systems with more complex dispersion relations?'. 
The current paper explores the important case where the dispersion is flat.
This research was supported in part by the National Science Foundation under Grant No. NSF PHY-1125915.
M. Valiente acknowledges support from EPSRC grant No. EP/M024636/1 and 
N.~T. Zinner acknowledges support from the Danish Council for Independent Research and the 
DFF Sapere Aude program.

\bibliographystyle{unsrt}

\begin{thebibliography}{99}

\bibitem{blume2012} D. Blume, Rep. Prog. Phys. {\bf 75}, 046401 (2012).
\bibitem{zinnerjensen2013} N.~T. Zinner and A.~S. Jensen,  J. Phys. G: Nucl. Part. Phys. {\bf 40} (2013) 053101 
\bibitem{zinner2014}  N.~T. Zinner, Few-Body Syst. Volume {\bf 55}, pp. 599-604 (2014).
\bibitem{naidon2016} P. Naidon and S. Endo, arXiv:1610.09805.
\bibitem{bkt1} V.~L. Berezinskii, JETP {\bf 32}, 493 (1971).
\bibitem{bkt2} V.~L. Berezinskii, JETP {\bf 34}, 610 (1972).
\bibitem{bkt3} J.~M. Kosterlitz and D.~J. Thouless, J. Phys. C {\bf 6}, 1181 (1973).

\bibitem{Peotta1} M. Tovmasyan, S. Peotta, P. T\"orm\"a and S. Huber, arXiv:1608.00976.
\bibitem{Peotta2}  A. Julku, S. Peotta, T.~I. Vanhala, D.-H. Kim and P. T\"orm\"a , Phys. Rev. Lett. {\bf 117}, 045303 (2016).
\bibitem{Peotta3} L. Liang, T.~I. Vanhala, S. Peotta, T. Siro, A. Harju and P. T\"orm\"a, arXiv:1610.01803.
\bibitem{Julien0} B. Dou\c{c}ot and J. Vidal, Phys. Rev. Lett. {\bf 88}, 227005 (2002).
\bibitem{Gershenson} S. Gladchenko {\it et al.}, Nature Phys. {\bf 5}, 48 (2009).

\bibitem{sutherland1986} B. Sutherland, Phys. Rev. B {\bf 34}, 5208 (1986).

\bibitem{Julien1} J. Vidal, R. Mosseri and B. Dou\c{c}ot, Phys. Rev. Lett. {\bf 81}, 5888 (1998).
\bibitem{Julien2} J. Vidal, P. Butaud, B. Dou\c{c}ot and R. Mosseri, Phys. Rev. B {\bf 64}, 155306 (2001).
\bibitem{Julien3} J. Vidal, B. Dou\c{c}ot, R. Mosseri and P. Butaud, Phys. Rev. Lett. {\bf 85}, 3906 (2000).

\bibitem{hyrkas2013} M. Hyrk{\"a}s, V. Apaja, and M. Manninen, Phys. Rev. A {\bf 87}, 023614 (2013).
\bibitem{zhang2015} T. Zhang and G.-B. Jo, Scientific Reports {\bf 5}, 16044 (2015).
\bibitem{phillips2015} L.~G. Phillips, G. De Chiara, P. {\"O}hberg, and M. Valiente, Phys. Rev. B {\bf 91}, 054103 (2015).
\bibitem{santos2004} L. Santos {\it et al.}, Phys. Rev. Lett. {\bf 93}, 030601 (2004).
\bibitem{yang2016} Z.-H. Yang {\it et al.}, Phys. Rev. A {\bf 93}, 062319 (2016).
\bibitem{morales2016} L. Morales-Inostroza and R.~A. Vicencio, Phys. Rev. A {\bf 94}, 043831 (2016).
\bibitem{mielke1991} A. Mielke, J. Phys. A: Math. Gen. {\bf 24}, 3311 (1991).
\bibitem{mielke1992} A. Mielke, J. Phys. A: Math. Gen. {\bf 25}, 4335 (1992).
\bibitem{mielke1993} A. Mielke and H. Tasaki, Commun. Math. Phys. {\bf 158}, 341 (1993).

\bibitem{baboux2016} F. Baboux {\it et al.}, Phys. Rev. Lett. {\bf 116}, 066402 (2016).
\bibitem{weimann2016} S. Weimann {\it et al.}, Optics Lett. {\bf 41}, 2414 (2016).
\bibitem{jo2012} G.-B. Jo {\it et al.}, Phys. Rev. Lett. {\bf 108}, 045305 (2012).
\bibitem{zong2016} Y. Zong {\it et al.}, Optics Exp. {\bf 24}, 8877 (2016).
\bibitem{vicencio2015} R.~A. Vicencio {\it et al.}, Phys. Rev. Lett. {\bf 114}, 245503 (2015).
\bibitem{taie2015} S. Taie {\it et al.}, Science Advances {\bf 1}, 1500854 (2015).
\bibitem{slot2016} M.~R. Slot {\it et al.}, arXiv:1611.04641.

\bibitem{wehling2014} T.~O. Wehling, A.~M. Black-Schaffer, and A.~V. Balatsky, Adv. Phys. {\bf 76}, 1 (2014).
\bibitem{sabio2010} J. Sabio, F. Sols, and F. Guinea, Phys. Rev. B {\bf 81}, 0454428 (2010).
\bibitem{huang2013} Z. Huang, D.~P. Arovas, and A.~V. Balatsky, New J. Phys. {\bf 15}, 123019 (2013).
\bibitem{gaul2014} C. Gaul, F. Dom{\'i}nguez-Adame, F. Sols, and I. Zapata, Phys. Rev. B {\bf 89}, 045420 (2014).
\bibitem{jiang2016} Q.-D. Jiang {\it et al.}, Phys. Rev. B {\bf 93}, 195165 (2016).
\bibitem{lothman2016} T. L{\"o}thman and A.~M. Black-Schaffer, arXiv:1611.04893.
\bibitem{chen2016} H. Chen and X. Niu, arXiv:1611.04975.

\bibitem{dalibard2011} J. Dalibard, F. Gerbier, G. Juzeliunas, and P. {\"O}hberg, Rev. Mod. Phys. {\bf 83}, 1523 (2011).
\bibitem{zhai2012} H. Zhai, Int. J. Mod. Phys. B {\bf 26}, 1230001 (2012).
\bibitem{galitski2013} V. Galitski and I.~B. Spielman, Nature {\bf 494}, 49 (2013).

\bibitem{mundo2013a} D. Maldonado-Mundo, L. He, P. {\"O}hberg, and M. Valiente, Phys. Rev. A {\bf 88}, 053609 (2013).
\bibitem{mundo2013b} D. Maldonado-Mundo, P. {\"O}hberg, and M. Valiente, J. Phys. B: At. Mol. Opt. Phys. {\bf 46} 134002 (2013).
\bibitem{wang2015} S.-J. Wang and C.~H. Greene, Phys. Rev. A {\bf 91}, 022706 (2015).
\bibitem{zhou2015} L. Zhou and X. Cui, Phys. Rev. B {\bf 92}, 140502(R) (2015).
\bibitem{guan2016} Q. Guan and D. Blume, Phys. Rev. A {\bf 94}, 022706 (2016).
\bibitem{wang2016} S.-J. Wang and C.~H. Greene, arXiv:1610.00198.


\bibitem{achilleos2013} V. Achilleos {\it et al.}, Europhysics Letters {\bf 103}, 20002 (2013).
\bibitem{marchukov2013} O. V. Marchukov {\it et al.}, J. Phys. B: At. Mol. Opt. Phys. {\bf 46} 134012 (2013).
\bibitem{guan2014} Q. Guan, X.~Y. Yin, S.~E. Gharashi, and D. Blume, J. Phys. B: At. Mol. Opt. Phys. {\bf 47}, 161001 (2014).
\bibitem{yin2014} X.~Y. Yin, S. Gopalakrishnan, D. Blume, Phys. Rev. A {\bf 89}, 033606 (2014).
\bibitem{marchukov2014} O. V. Marchukov {\it et al.}, J. Phys. B: At. Mol. Opt. Phys. {\bf 47} 195303 (2014).
\bibitem{guan2015} Q. Guan and D. Blume, Phys. Rev. A {\bf 92}, 023641 (2015).

\bibitem{shi2014} Z.-Y. Shi, X. Cui, and H. Zhai, Phys. Rev. Lett. {\bf 112}, 013201 (2014).
\bibitem{cui2014} X. Cui and W. Yi, Phys. Rev. X {\bf 4}, 031026 (2014).
\bibitem{shi2015} Z.-Y. Shi, H. Zhai, and X. Cui, Phys. Rev. A {\bf 91}, 023618 (2015).

\bibitem{Phillips} M. Valiente, L.~G. Phillips, N.~T. Zinner and P. \"Ohberg, arXiv:1505.03519.

\bibitem{huber2010} S.~D. Huber and E. Altman, Phys. Rev. B {\bf 82}, 184502 (2010).

\bibitem{Tomonaga}
S. Tomonaga, Progr. Theoret. Phys. (Kyoto) {\bf 5}, 544 (1950).

\bibitem{Luttinger}
J.~M. Luttinger, J. Math. Phys. {\bf 4}, 1154 (1963).

\bibitem{MattisLieb}
D.~C. Mattis and E.~H. Lieb, J. Math. Phys. {\bf 6}, 304 (1965).

\bibitem{MattisSutherland}
D.~C. Mattis and B. Sutherland, J. Math. Phys. {\bf 22}, 1692 (1981).


\end{thebibliography}

\end{document}